\documentstyle[prl,aps,multicol]{revtex}

\newcommand{\scri}{{\cal I}}
\newcommand{\pd}{\partial}
\newcommand{\E}{{\cal E}}
\newcommand{\rr}{{\bf k}}
\newcommand{\rT}{{R^{n-3}/\Gamma}}
\newcommand{\rs}{{\bf  r}_{s {\rm +}}}
\newcommand{\rb}{{\bf  r}_{b {\rm +}}}
\newcommand{\ra}{{\bf  r}_{\alpha{\rm +}}}
\newcommand{\M}{{\cal M}}
\newcommand{\cB}{{\cal B}}
\newcommand{\cA}{{\cal A}}
\newcommand{\ft}{{flat\ }}
\newcommand{\Ft}{{Flat\ }}
\newcommand{\kk}{{\kappa}}
\newcommand{\tI}{{\widetilde I}}

\begin{document}

\title{Phase Transitions for \Ft adS  Black Holes}
\author{Sumati Surya$^\ast$, Kristin Schleich$^\dag$, Donald 
M. Witt$^\ddag$} 
\address{Department of Physics and Astronomy, 
University of British Columbia,\\
6224, Agricultural Road, Vancouver, 
B.C., V6T 1Z1, Canada \\ 
$^\ast$ssurya@pims.math.ca, $\,\,$
$^\dag$schleich@noether.physics.ubc.ca,  
 $\,\,$  $^\ddag$donwitt@noether.physics.ubc.ca }
\date{\today}

\maketitle
\begin{abstract}
{ We reexamine the thermodynamics of adS black holes with Ricci flat
horizons using the adS soliton as the thermal background. We find that
there is a phase transition which is dependent not only on the
temperature, but also on the black hole area, which is an independent
parameter. As in the spherical adS black hole, this phase transition
is related via the adS/CFT correspondence to a
confinement-deconfinement transition in the large $N$ gauge theory on
the conformal boundary at infinity. }

\end{abstract}

\begin{multicols}{2} 

Black hole thermodynamics in anti-de Sitter (adS) spacetimes is of
current interest through its connection to large N gauge theories. A
study of the thermodynamics of spherically symmetric adS black holes
by Hawking and Page \cite{hawpage} showed that there is a phase
transition: high temperature black holes are stable while low
temperature black holes are unstable and decay to the global adS
spacetime. Using the adS/CFT correspondence \cite{maldacena},
it was shown by Witten \cite{witten,witten98} that in the large N
limit of an ${\cal {N}}=4 \,\,$ SU(N) gauge theory on the conformal
boundary, this phase transition corresponds to a
deconfinement-confinement transition. Whether or not
 such transitions occur for black holes with horizons of more
complicated topology is therefore of interest since it can provide
insights into the large N limit of gauge theories on different
topologies.

Topological black holes in asymptotically adS spacetimes were first
found by \cite{lemos95,tgbhs} in three and four
spacetime dimensions (see also \cite{vanzo}). These solutions have
been generalised to arbitrary dimensions in
\cite{banados98,mann,dbir} where the black hole horizons are
Einstein spaces of positive, negative or zero curvature.  Being
static, they lend themselves to the standard equilibrium
thermodynamics analysis \cite{vanzo,dbir,blp,emparan}. It was shown in
\cite{vanzo,dbir} that there are no phase transitions for adS black
holes with Ricci flat horizons when the zero mass black hole is taken
as the thermal background.

We show that the choice of background is crucial: using the adS
soliton of Horowitz and Myers \cite{hm} as the thermal background
instead, we find that there is indeed a phase transition for adS black
holes with Ricci flat horizons.  This choice is a natural one, as the
adS soliton is conjectured to be the lowest energy solution for
spacetimes which, like adS black holes with Ricci flat horizons,
asymptotically approach the adS soliton sufficiently rapidly
\cite{hm}.  We show that, in 4 or more dimensions,  
there are two independent parameters that
determine this phase transition: the temperature and the black hole
area.  This behavior differs from that of the spherical adS black
holes in 4 or more dimensions for which the phase transition depends only on the temperature.
(In 3 dimensions, the phase transition is identical to the spherical
case, since the  adS black hole with Ricci flat horizon is the same
as the spherical adS black hole, and the adS soliton is diffeomorphic to global adS.)
Following \cite{witten,witten98} we argue that the deconfinement phase
corresponds to a large, cold black hole, while the confinement phase
corresponds to a small, hot adS soliton. As in the spherical case, the
deconfinement-confinement transition can be seen to occur only for
anti-periodic spinors. 


A useful form of the $n$ dimensional adS soliton of \cite{hm} for
negative cosmological constant $\Lambda$   is
\begin{equation}
ds_{s}^2 = -r^2dt_s^2 + \frac{dr^2}{V_s(r)} + V_s(r)  d\phi_s^2 + r^2
h_{ij}(\Theta)d\theta_id\theta_j\  \label{soliton.eq}
\end{equation} 
where 
\begin{equation}  
\label{function.eq}
V_s(r)=\biggl(\frac{r^2}{l^2}- \frac{\rr_s^{n-1}}{r^{n-3}}\biggr),\,\, \Lambda =
-\frac{(n-1)(n-2)}{2 l^2}\ . \end{equation} 
Here, $h_{ij}(\Theta)$ is a Ricci flat metric on the quotient manifold
$\rT$ where $\Gamma$ is a finite discrete group  and  the
$\Theta=\{\theta_i\}$ are coordinates on this manifold.
A simple example of such an $\rT$ is the torus $T^{n-3}$. 
Regularity requires $r\geq \rs$, where $V_s(\rs)=0$, and
$\phi_s$ to be identified with period $\beta_s=\frac{4\pi
l^2}{(n-1)\rs}$. The adS soliton has a flat conformal boundary
$\scri_s$ of topology $R\times S^1 \times \rT$, as opposed to
global adS which has a positive curvature $\scri$ of topology $R
\times S^{n-2}$.

Horowitz and Myers conjectured that in five dimensions any spacetime
which asymptotically approaches the soliton metric ${\bar g}_{\mu
\nu}$, i.e., $g_{\mu\nu}= {\bar g}_{\mu \nu} + h_{\mu\nu}$, such that
$h_{\gamma\rho}=O(r^{-2}), \, h_{\gamma r}=O(r^{-4})$ and
$h_{rr}=O(r^{-6})$, for $ \gamma, \rho \neq r$, must have energy $E
\geq 0$ with respect to the soliton, the equality being satisfied only
by the soliton itself.  This conjecture is therefore restricted to
those spacetimes whose $\scri$ and asymptotics both match that of the
soliton.

The  metric of adS black holes with Ricci flat horizons (henceforth called
flat adS black holes) can be written 
\begin{equation} 
\label{zero.eq}
ds_{bh}^2 = -V_b(r)dt_b^2 + \frac {dr^2}{V_b(r)} 
+ r^2 d\phi_b^2 +r^2 h_{ij}(\Theta)d\theta_id\theta_j 
\end{equation} 
where $V_b(r)$ is the function (\ref{function.eq}) with the parameter
$\rr_b$ replacing $\rr_s$.  When $\phi_b$ is identified with period
$\eta_b$, this solution has a conformal boundary $\scri_b \approx R
\times S^1 \times \rT$ which is diffeomorphic to that of the soliton,
$\scri_s$.  Note that the horizon is at $V_b(\rb)=0$, and has the same
topology as $\scri_b$.  The zeros of $V_{\alpha}(r)$ are given by
$\ra^{n-1}=\rr_{\alpha}^{n-1}l^2$, where we have introduced the label
$\alpha \in \{b,s\}$ for convenience.  The area of the black hole
horizon for $\rT$ compact is,
\begin{equation} 
\label{area.eq}
A_b = {\rm vol}(\rT)\eta_b  \rb^{n-2}.       
\end{equation}   
The black hole metric (\ref{zero.eq}) asymptotically approaches the
adS soliton with the required $n$-dimensional generalisation of the fall
off conditions of \cite{hm}, so that the soliton can indeed be taken as
the background. 
We note that for $n=3$, the flat adS black hole and the spherical adS
black hole are the same and that the $n=3$ soliton is in fact just
global adS. Since the spherical case is well understood, we will
restrict our analysis to $n>3$.

We now calculate the energy of the black hole with respect to the
soliton background using the standard regularisation scheme, in which
 we match the two solutions at some finite cut off radius
$R$, calculate the energy as a function of $R$, then take the
limit, $R \rightarrow \infty$. The matching condition at finite $R$
is,
\begin{equation}
\label{ematch.eq}
\beta_b{\sqrt{V_b(R)}}=R\eta_s,
\end{equation}    
along with the requirement that the metrics on $\rT$ match   
for both solutions. In the special case $\rT \approx {\rm T}^{n-3}$, this means
that the periodicities of the $\theta_i$ are the same for the soliton
and the black hole.   The regulated energy of the black hole is,  
\begin{equation}
\E(R)= -\frac{1 }{ 8 \pi G} \int d\mu N_b(K_b -K_s),  
\end{equation}  
where $N_b$ is the lapse for the black hole, the $K_\alpha$ are
extrinsic curvatures of the ``matching'' boundary at $R$ for the two
solutions and $d\mu$ is the volume element of the surface at $R$. 
 Using (\ref{ematch.eq}) and taking $R \rightarrow \infty$,
we get
\begin{equation}
\label{energy.eq}
\E= \frac{ {\rm vol}(\rT)}{ 4 G}\frac{l}{(n-1) \rs }
[(n-2)\rr_b^{n-1} +\rr_s^{n-1}],  
\end{equation}  
which is always positive with respect to the soliton, in keeping with
the Horowitz-Myers conjecture.

In order to study the thermodynamics of the black hole in the
semiclassical limit, we  go to the Euclidean sector
by a Wick rotation $t_\alpha \rightarrow i\tau_\alpha$, 
\begin{equation} 
ds_{b}^2=V_b(r)d\tau_b^2 + \frac{dr^2}{V_b(r)} + r^2 d\phi_b^2 +r^2
h_{ij}(\Theta)d\theta_id\theta_j, 
\label{ebh.eq}
\end{equation} 
where regularity demands that $\tau_b$ be identified
with period, $\beta_b=\frac{4\pi l^2}{ (n-1)\rb}$. The Euclideanised
soliton is, 
\begin{equation} 
ds_{s}^2= r^2d\tau_s^2 + \frac{dr^2}{V_s(r)} + V_s(r) d\phi_s^2 
+r^2 h_{ij}(\Theta)d\theta_id\theta_j,  
\label{esol.eq} 
\end{equation} 
where $\tau_s$ can have an arbitrary periodicity $\eta_s$.

Next, we calculate the action of the black hole (\ref{ebh.eq}) of mass
parameter $\rr_b$ in the background of a soliton (\ref{esol.eq}) of mass
parameter $\rr_s$.  Again, as the actions are infinite, we employ  the
regularisation procedure used to calculate the energy
(\ref{energy.eq}). The matching conditions include 
\begin{equation} 
\label{bsmatch.eq}
\beta_b{\sqrt{V_b(R)}}=R\eta_s,  \quad
\beta_s{\sqrt{V_s(R)}}=R\eta_b, 
\end{equation} 
and the matching of metrics on $\rT$.  Although (\ref{bsmatch.eq})
depends on $R$, the relations between the periodicities is only
meaningful in the limit $R \rightarrow \infty$ in which $\beta_b =
\eta_s l,\ \beta_s = \eta_bl$, and are independent of $R$.  It is
clear that (\ref{bsmatch.eq}) can be satisfied in this limit even when
$\rr_b \neq \rr_s$.

The Euclidean action for a vacuum solution of Einstein's equations
with cosmological constant
on a manifold $\M$ with boundary $\pd \M$ is,
\begin{eqnarray} 
\tI &=& -\frac{1}{16 \pi G}\int_\M \!\!\!d^nx{\sqrt{g}}(R-2\Lambda)\!
-\frac{1}{8 \pi G}\int_{\pd\M} \!\!\!d^{n-1}x \sqrt{h}K \nonumber \\ 
& =& \frac{(n-1)}{8 \pi G l^2}\int_\M d^nx {\sqrt g} -\frac{1}{8 \pi G}\int_{\pd\M}d^{n-1}x \sqrt{h} K. 
\end{eqnarray} 
In terms of the cut-off radius $R$, the regularised black hole action is, 
\begin{eqnarray} 
I(R) &=& \tI_{b}(R) - \tI_{s}(R)=\frac {(n-1)\, {\rm vol}(\rT)}{ 8 \pi
G l^2} \times \nonumber \\ 
&& \Biggl[ \beta_b\eta_b \int_{\rb}^R r^{n-2} dr -\beta_s\eta_s
\int_{\rs}^R r^{n-2} dr\Biggr],
\end{eqnarray} 
where we have ignored the difference in boundary terms, since this
vanishes in the large $R$ limit.  Using the matching conditions
(\ref{bsmatch.eq}) and taking the regulator to infinity we get,
\begin{eqnarray} 
I = \lim_{R \rightarrow \infty} I(R) &=& \frac{{\rm vol}(\rT)}{16
\pi Gl^3} \biggl(\frac{4 \pi l^2}{n-1}\biggr)^{n-1} \times \nonumber
\\
&& \beta_b\beta_s [\beta_s^{1-n} -\beta_b^{1-n}]. 
\end{eqnarray} 
We see that the action and hence the free energy $F= \beta_b^{-1} I$,
is less than zero when $\rr_b > \rr_s$ and is greater than 
zero for $\rr_b<\rr_s$,  signalling a phase transition.
Since the soliton is the conjectured lowest energy state, we take its
free energy to be identically zero.

This phase transition might appear to be exactly analogous to the
Hawking-Page transition for spherical adS black holes \cite{hawpage},
which has a phase structure dependent only on temperature: high
temperature black holes are stable while low temperature black holes
decay to the global adS spacetime.  However, there are important
distinctions, which can be traced partly to the fact that the area and
the temperature of the \ft adS black holes are independent quantities,
while they are not for the spherical adS black hole.

Writing the area of the black hole (\ref{area.eq}) in terms of the
$\beta_\alpha$'s,
\begin{equation} 
A_b = \kk \beta_b^{(2-n)} \beta_s, \,\, \kk= \frac{{\rm vol}(\rT)}{l}\biggl(\frac{4\pi l^2
}{n-1}\biggr)^{n-2}  \end{equation}
we find that the ratio, 
\begin{equation}
\frac{A_b }{ \beta_b^{3-n}}= \kk \frac {\rr_b }{ \rr_s},        
\end{equation} 
helps us characterise the phase transition.  For $n>3$, if we
normalise ${\rm vol}({\rT})$ so that $\kk=1$, we see that small, hot
black holes, i.e., $A_b << \beta_b^{3-n}$ or $\rr_b << \rr_s$, are
unstable and decay to small, hot solitons. On the other hand, large,
cold black holes, i.e., $A_b >> \beta_b^{3-n}$ or $\rr_b >>\rr_s$ are
stable. Black holes with $A_b \sim \beta_b^{3-n}$ are thus in
equilibrium with the soliton and include the cases when they are large
and hot or cold and small.  Thus, the stability of the \ft black hole
is not merely dependent on the temperature of the black hole but also
on its size.

Now, if we had taken $\rr_b=\rr_s$, and matched the periodicities of
$\tau_\alpha$ and $\phi_\alpha$ appropriately, we would have found the
two instantons to be identical and consequently there would be no
phase transition at all. However, constraining ourselves to this
special case obviously trivialises the thermodynamics of the black
hole in the background of the soliton, since it represents only the
critical point in the phase diagram.

Note that the thermodynamics of the spherical adS
black hole is substantially different from that of the asymptotically
flat case, namely the Schwarzschild solution, as was first pointed out
in \cite{hawpage}. In particular, while Schwarzschild black holes have
negative specific heat, sufficiently massive spherical adS black hole
have positive specific heat. This makes the canonical ensemble for the
adS black holes perfectly valid and does not require the use of an
artificial box to ensure equilibrium conditions. However, the specific
heat does have a first order pole at a critical mass, below which it
is negative and hence unstable.

The \ft  adS black hole in the background of a soliton is quite
different in this respect.  The thermodynamic energy, $E=\frac{\pd I
}{ \pd \beta_b}$,  
\begin{eqnarray} 
E &=&  \frac{{\rm vol}(\rT) }{16 \pi
G l^3} \biggl(\frac{4\pi l^2 }{ n-1}\biggr)^{n-1} \beta_s 
[(n-2)\beta_b^{1-n}+\beta_s^{1-n}] \nonumber\\  
&=& \frac{{\rm vol}(\rT) }{16 \pi
G l} \beta_s [(n-2)\rr_b^{n-1} +\rr_s^{n-1}] ,   
\end{eqnarray} 
matches the energy calculated in (\ref{energy.eq}) and increases
monotonically with the temperature of the black hole.  But, $\frac{\pd
E }{ \pd A_b}<0$ which implies that as the area of the horizon
increases, the energy decreases.

The specific heat is, 
\begin{equation} 
C = \frac{\pd E }{ \pd T_b}=\frac{ (n-2) }{ 4 G} A_b,
\end{equation} 
which, being proportional to the area is {\it always} positive.  It
is, moreover, independent of the temperature of the black hole: 
larger black holes take more energy to heat up irrespective of what
temperature they are at, in comparison with smaller black holes. We
note that in \cite{dbir} the calculation of the specific heat using
the zero mass black hole as background also gives a positive specific
heat.

Finally, a calculation of the entropy yields, 
\begin{equation}  
S=\beta_b \frac {\pd I }{ \pd \beta_b}= \frac{A_b  }{ 4  G},   
\end{equation} 
as expected. 


A version of the adS/CFT correspondence tells us that the partition
function for a conformal field theory on a manifold $\cB$ is
equivalent to the sum of partition functions of string theory over all
manifolds $\M$ with conformal boundary $\pd \M=\cB$
\cite{witten,witten98}, which further satisfy certain boundary
conditions on the fields. For example for $n=5$, the appropriate
boundary conformal field theory is ${\cal{ N}}=4$ SU(N) gauge theory.

The correspondence tells us that the large $N$ limit of the
field  theory is equivalent to the semiclassical limit of the string
theory. In this limit,  the  string partition function becomes
simply a sum over all negative curvature Einstein manifolds with boundary
$\cB$, so that,  
\begin{equation}
\label{part.eq}
\lim_{N\rightarrow \infty} Z_{CFT}(\cB)=\sum_{\stackrel{\M}{\pd
\M=\cB}} \exp^{-I(\M)}.   
\end{equation} 
Thus the semiclassical black hole phase transition must correspond to
a phase transition in the large $N$ limit of the gauge theory
\cite{witten,witten98}.  
For the spherical adS black hole  the Hawking-Page phase transition 
was interpreted by \cite{witten,witten98} to correspond to a
deconfinement-confinement transition in the gauge theory using an
essentially topological argument. We see here that these arguments can
be trivially extended to the \ft  adS black holes as well.

As we have seen, for a conformal boundary $\cB \approx S^1 \times S^1
\times \rT$, there are contributions to the partition function from
both the flat adS black hole and the adS soliton, so that the string
partition function includes in the sum these two Einstein
manifolds.

Both the black hole and the soliton have topology $B^2 \times S^1
\times \rT$. What distinguishes the two phases in the gauge theory,
however, is the fact that the expectation value of the temporal Wilson
loop operator $W(C)$ is distinct in each case. Now, $<W(C)>$ is an
order parameter for the spontaneous symmetry breaking of a subgroup of
the center of the gauge group corresponding to the deconfinement
phase. In the large $N$ limit, $<W(C)>$ is essentially proportional to
$\exp^{-\cA(D)}$, where $D$ is the string world sheet with boundary
$C$ which has the smallest regularised world volume, $\cA(D)$. For the
soliton, the temporal loop $C$ must lie along the $S^1$ factor which
is non-contractible. Since there can be no world sheet $D$ with
boundary $\pd D= C$, $<W(C)>$ must vanish.  For the black hole, the
temporal loop $C$ must lie in the $B^2$ factor and is thus just the
boundary of the disc $D=B^2$, which implies that $<W(C)>$ is
non-zero. The stable black hole phase $\rr_b>\rr_s$ therefore
corresponds to deconfinement whereas the stable soliton phase
$\rr_b<\rr_s$ corresponds to confinement. The so called ``singular''
points of the gauge theory correspond to the case $\rr_1 \sim \rr_2$.
Note that \cite{kalyanbala} obtained this  point for boundary
topologies which include an $S^1\times S^1$ factor,  using   
qualitative arguments  on the form of the gauge theory partition function. 
However, our identification of the adS soliton as the background
allows us to calculate this explicitly. 

For the spherical adS black hole, it was argued in \cite{witten} that
a phase transition can only occur for the anti-periodic partition
function, ${\rm Tr}\exp^{-\beta H}$.  This argument,
which  depends on the spin structure in the temporal direction alone,
has a trivial extension to the \ft adS black hole: while the soliton
has a temporal $S^1$ which is non-contractible and can support both
anti-periodic and periodic spinors, the temporal $S^1$ of the \ft adS
black hole is contractible and can support only anti-periodic
spinors. Thus, although both the soliton and the black hole have the
same topology, the confinement-deconfinement phase transition can occur
only for the anti-periodic case.

From the gravity point of view, the deconfinement phase for the \ft
case differs from that of the spherically symmetric one. In the
latter, this phase is represented by a black hole at high temperature
which could be either very small or very large. This is because the
black hole radius is double valued with respect to the temperature. On
the other hand, for the $n>3$ \ft adS black hole, the deconfinement
phase is represented by a black hole whose area is much greater than
its temperature, i.e.  by large cold black holes. If we fix the size
of the black hole, we see that the deconfinement phase occurs at low
temperatures instead of high temperatures.

As pointed out in \cite{witten98}, it is the ratio of the inverse
temperature to the ``size'' of the spatial section on the conformal
boundary which is relevant for the field theory. Of particular
interest is the ``infinite volume'' limit where the temperature is
kept fixed while the spatial volume is taken to infinity. What is
meant by this is a ``limit'' in which the spatial geometry approaches
that of a flat $R^{n-2}$.  For the spherical case, there is only one
free parameter, the temperature, and hence it is not apriori clear
what one is varying in taking the limit.  As we have seen, for the \ft
case there are two parameters: the temperature and the periodicity
$\eta_b$ of the spatial $S^1$ factor, so that this limit appears more
natural. Specifically, for the case of a black hole spacetime with
trivial $\Gamma$, i.e., $R^{n-3}/\Gamma = R^{n-3}$, one takes $\eta_b
\rightarrow \infty$, so that the geometry of the spatial section on
the conformal boundary approaches flat $R^{n-2}$.  In this limit,
$\rr_b>>\rr_s$ and so there is only one phase, the deconfining phase,
as expected from the analysis in \cite{witten98}.


Finally, in the absence of a proof of the Horowitz-Myers conjecture,
the choice of the adS soliton as the preferred thermal background is
conjectural; it is possible that a spacetime of lower mass than the
soliton exists. However, it is clear from our results that {\it some}
phase transition must exist. This is clearly linked to the larger
problem of whether or not there is a positive energy theorem for
asymptotically adS spacetimes which yields nonsingular minimum energy
backgrounds for any given topology of $\scri$. In the special case
when $\scri$ has topology $S^{n-2} \times R$, arguments for the
boundedness of mass \cite{abdes,ew} indicate that a zero energy
solution is the appropriate background. Thus the choice of global adS
seems physically reasonable, putting the Hawking Page transition on
firm footing.  We note that no candidate non-singular background analogous
to the adS soliton seems to exist for the hyperbolic black holes,
which leaves their semiclassical thermodynamics relative to such a
background still unexplored.


After this work was completed, we were informed that the possibility
that such phase transitions might occur for toroidal boundaries was
suggested in a footnote in \cite{stringphase}.  We thank R. Myers and
G.  Horowitz for useful comments on the first version of the paper.
S.S. would like to thank E. Woolgar and K. Zarembo for extensive
discussions.  S.S. and K.S. acknowledge support from National Science
and Engineering Research Council of Canada. S.S. also thanks the
Pacific Institute for the Mathematical Sciences for support.

\end{multicols}
\end{document}